\documentclass[runningheads]{llncs}
\usepackage[T1]{fontenc}
\usepackage{graphicx}
\usepackage{hyperref}

\def\simpropto{\lower.2ex\hbox{$\; \buildrel \propto \over \sim \;$}}
\def\ltsim{\lower.5ex\hbox{$\; \buildrel < \over \sim \;$}}
\def\gtsim{\lower.5ex\hbox{$\; \buildrel > \over \sim \;$}}

\begin{document}
\title{The Limits of Cosmology}
%
%
\author{Joseph Silk}


%
\authorrunning{Silk}
%
\institute{{Institut d' 
Astrophysique de Paris, UMR 7095 CNRS \& Sorbonne Universit\'{e}, 98bis Boulevard Arago, F-75014 Paris, France}
\and{William H. Miller III Department of Physics and Astronomy, The Johns Hopkins University, 3400 N. Charles	Street, Baltimore, MD 21218, U.S.A.}
\and{Beecroft Institute for Particle Astrophysics and Cosmology, University of Oxford, Keble	Road, Oxford OX1 3RH, U.K.}\\
\email{silk@iap.fr}}

\maketitle              
\begin{abstract}
The Moon is our future. It may seem like a chimera with a projected cost in excess of 100 billion\$, and counting, dispensed on ARTEMIS with little to show to date. However
 it is the ideal site for the largest telescopes that we can dream about,  at wavelengths spanning decimetric radio through optical to terahertz FIR.
And it  is these future telescopes  that will penetrate the fundamental mysteries of the first hydrogen clouds, the first stars, the first galaxies, the first supermassive black holes, and  the nearest habitable exoplanets. Nor does it stop there. Our lunar telescopes will take us back to the first months of the Universe, and even back to the first 10$^{-36}$ second after the Big Bang when inflation most likely occurred.  Our lunar telescopes will provide high resolution images of exoplanets that are nearby  Earth-like  'twins'  and provide an unrivalled attempt to answer the ultimate cosmic question of whether we are alone in the universe.  Here I will set out my vision of the case for lunar astronomy over  the next several decades. 

\end{abstract}
\section {Half a century ago}
The Apollo era was launched  with an iconic speech by US President John F. Kennedy in 1962. It was his response to the completely unexpected  launch of Sputnik, the first space satellite, by the Soviet Union.

These historic words echoed around the world.
{\it We choose to go to the Moon! We choose to go to the Moon...We choose to go to the Moon in this decade and do the other things, not because they are easy, but because they are hard; because that goal will serve to organize and measure the best of our energies and skills, because that challenge is one that we are willing to accept, one we are unwilling to postpone, and one we intend to win.} The space race had begun in earnest.

The Apollo project was  hugely successful. Some 12 US astronauts walked on the Moon. But it was a short-lived project, simply  too expensive to maintain. There was no  lunar sequel, in   part due to the huge costs but equally because there was little competition after the fall of the former Soviet Union.  But exploration of the Universe from space  was to begin in earnest. The next decades saw the development of the International Space Station, the space shuttle, and the launch of the first space telescopes, along with international  space agencies playing a major role.

Only in recent years has the situation with regard to the potential of crewed lunar exploration changed,  with the emergence of China as a space powerhouse.  It took half a century, but now we are back in the race to the Moon. The new space race will do more than install astronauts on the Moon. Much more is envisaged.  Apart from  the commercial exploitation of lunar resources, including rare minerals and ice mining for fuel, goals include construction of orbiting lunar space stations, with science projects playing a central role.
The science envisaged will include new generations of lunar telescopes that can accomplish targets that are not  attainable from  the earth or even in space. I will describe some of these below.

One common feature of the experiments  that I will describe is that the science is unique and compelling, and will explore new frontiers in cosmology and astronomy, spanning the very early universe  to astrometry and imaging of  nearby exoplanets. Because of the unique properties of the Moon  that are relevant for astronomy, most notably the lack of any atmosphere or ionosphere, and the seismologically quiet conditions of the lunar surface, we will. be able to build telescopes that will achieve science  that cannot be realized from the Earth or even via space satellites. Such telescopes will provide a unique opportunity to probe beyond  the current  limits of cosmology and even of exoplanet research. 
   
\section {A century ago}
Modern cosmology effectively began with Georges Lema\^{i}tre in 1927. He developed the cosmological constant as an inevitable ingredient in the expanding universe  solution of Einstein's  general relativity equations. This   led to a sequence of possible expanding universes  with a cosmological constant that previewed stalling or  more generally acceleration at late (or even early) times. Perhaps his greatest achievement was to 
bring  the new physics into cosmology by incorporating the concept of vacuum energy fluctuations as the physics interpretation  of the cosmological constant term.  Writing during the period soon after  the birth of quantum physics that replaced Bohr's theory of the atom, 
he asserted that {\it we must assign a pressure $p=-\rho c^2$ to the density energy $\rho c^2$. This is essentially the meaning of the cosmological constant $\lambda$ which corresponds to a negative density of vacuum.}

There was no looking back. A new generation of physicists was inspired  to enter cosmology  half a century later to develop the theory of inflation, designed to account for such essential features as the size of the observable universe, the Euclidean nature of the universe, and the quantum origin of the primordial density fluctuations that seeded structure formation.

\section {The biggest questions in physics}
We are at a turning point in cosmology. There are certain crisis points in cosmology where we are limited by data. These have led to growing tensions in terms of understanding the Universe, most  notably that of the rate of expansion of the Universe.   Equally  critical for our current best model  of cosmology  is that of understanding how the Universe began.

We need to find the best way forward.  Only larger telescopes will help us resolve current tensions. To see the  first galaxies and stars, we need really large telescopes. We have access to very large terrestrial telescopes, around 8-10m, and soon to be 39m with anticipated first light from the ELT in 2029. But that is  surely the limiting aperture  for optical or infrared telescopes on Earth.  
We have  succeeded over the past two decades in greatly reducing the error bars on the cosmological constant, aka dark energy. But we very far from finding deviations from the canonical Lema\^{i}tre value.   All recent experiments that have greatly improved accuracy are converging on Lema\^{i}tre's constant value at the current epoch. There is no guarantee that increased precision will take us into new physics territory.

There are recent hints of time variation in dark energy, but again, the systematic errors are so challenging that  it seems  unlikely that experiments such as DESI will take us unambiguously into the realm of new physics. New and larger experiments may well be needed. But without more convincing guarantees of success, compelling arguments will be needed for such costly endeavours  to be sufficiently competitive to be built over the next decades in the absence of stronger  theoretical motivation.

CMB fluctuation studies have a long and glorious history in cosmology, largely establishing the era of precision cosmology where we debate error bars on cosmological parameters of 1-2\%. The ultimate goal of the next generation of CMB studies is detection of the tensor component of primordial fluctuations that are of quantum origin and seeded the large-scale structure of the Universe. The experimental goal of experiments such as SO and LiteBIRD is detection of the primordial tensor mode, This would be a smoking gun for inflation. However only for the simplest classes  of inflation models, notably for single field slow roll inflation, is there any significant possibility of detection above the likely experimental threshold of $\sigma(T/S) \sim 0.001$ \cite{kamionkowski2016}. Generic multifield inflation models 
 have no robust lower bound on an inflationary gravity wave signal, in large part because of post-inflationary reheating uncertainties \cite{hotinli2018}. 

In space,  we are not limited by our atmosphere and have a clear view of the distant universe. Our largest telescope has a 8m aperture, the James Webb Space telescope.   To probe the earliest epoch of the universe, telescope  aperture is a critical factor as it controls light gathering power. It is  unlikely we will ever build a larger free flyer, it is too prohibitive in cost.  Hence we are telescope aperture-limited.  While  the relatively small aperture of JWST is ideal for probing the most distant galaxies to $z\sim 15, $  a new generation of wide aperture,  smaller diameter, telescopes is revolutionising astronomical surveys. Space telescopes such as EUCLID and ROMAN will generate surveys of billions of galaxies to $z\sim 1$ or beyond.  These surveys will refine  our precision in determining the parameters of our standard cosmological model.  

More excitingly perhaps, they will explore the boundaries of our standard model by seeking any possible variations, spatially or temporally, in the parameters that describe our local universe. However  our current model gives an exquisite fit to our local universe, with only very modest rumblings of discontent in,   for  example,  the degree of robustness in possible expansion rate or dark energy variations. Here the systematic errors in comparing diverse datasets are highly disputed, so no significant conclusions can yet be extracted, either with present or even most likely via forecasts from projected future data sets. 

Dark matter detection signals via multiton detector targets or indirect  signals via gamma ray astronomy remain elusive. One  really  has to choose future experiments that offer significant upgrades, for example  with higher resolution x-ray or gamma ray satellites. Yet there is no guarantee we will ever detect a dark matter signal. The potential parameter space is vast. Given our limited resources, one ideally would like to see next generation experiments with a guaranteed science return, as is the case, for example, in the area of gravitational wave astronomy.   In this area, our    most optimistic hope is that dark sirens may resolve  one of the major tensions in cosmology, that of the expansion rate of the universe. 
Even here,  the likely systematics are unknown.

The challenges go well beyond cosmology. Consider the exponentially expanding field of exoplanet research.  The ultimate goal is to find evidence of life in any form in distant exoplanets. We seek any of  several potential signatures  in the atmospheres or on the surfaces of earth "twins",  with rocky cores and in the habitable zones of  nearby F or G stars. Yet even a pioneering  experiment such as  NASA's  multibillion dollar  Habitable Worlds Observatory,  a 6m  space telescope with a precision coronagraph that is not scheduled to launch until the 2040s at the earliest,  will study in exquisite detail only a handful of the nearest Earth-like exoplanets. We anticipate a very limited number of targets in the form of potentially life-hosting exoplanets around nearby stars. Some 25  were in the project horizon  at last count but these numbers seem certain to decrease as descoping inevitably occurs because of cost overruns. 

Given the extreme uncertainties in quantifying the predictions of life developing in any form, the chances of a successful detection of robust biosignatures seem very low.
Similar comments may be made on the prospects of ongoing and proposed searches closer to home, such as in the sub-glacial oceans of Europa  or Enciladus,  or even via deep sample return from Mars, one that I find the most promising of proposals to date because of the early oxygen-rich atmosphere that characterized  young Mars some billions of years ago
 
 In high energy particle physics, the primary goal is to build particle accelerators that take us beyond the standard model. We know that the standard model of particle physics  cannot be the final word. To advance, we need to go beyond the 10TeV reach of the Large Hadron Collider  in searching the debris of particle collisions for clues such as detection of the weakly interacting particles that almost certainly  constitute dark matter.  A future circular collider that achieves  100 TeV rest frame particle collision  energies is likely to be 
 the next frontier in high energy physics. Design and construction will take at least 40 years and the cost could be up to 10 times   as much as the cost of the LHC.

Nevertheless this megaproject seems  likely to go ahead as  China, Europe and the US have such next generation large particle, muon or proton,  colliders in their planning horizons. However while such  experiments will tell us much about the particles that we know exist, there is no guarantee of the discovery of  new particles such as the elusive DM candidates or of exploration pf the long anticipated  paradise of SUSY.

There is a common thread running through all of these highly worthwhile experiments (Figure 1). Most  current large projects in astronomy  and in particle physics face the following dilemma. We can  significantly improve current limits in the future, but given finite  resources we need to be selective. Ideally we would like to base our selection criteria on the prospects of a major science advance.  The fundamental problem is that for essentially all future experiments, there is no guarantee of success  however much funding we have at our disposal.  These are generally experiments that most physicists think we must do, but we will undoubtedly have to make choices.  Criteria are needed that are usually ranked by  the prospects for compelling science given the vast budgets anticipated for essentially all of these next generation projects.  

However I believe that even this approach is not good enough when we face the daunting prospects of such costly endeavours. Humanity has even more compelling issues that need to be addressed. The budget for science is especially limited by this competition. 

However there is one area where the budget is not a completely dominant issue, and indeed is essentially committed,  and indeed when commercial interests are included, essentially unlimited. A key corollary is that the potential science return is not merely compelling but guaranteed. That is, from the astronomical perspective, we can look forward to  the installation of major telescope projects on the Moon. We are returning to the Moon in a massive way, spurred on by international competition.  Its a race initially for exploration and inevitably for exploitation that we (insert the US, Europe, China, India, Russia for starters) cannot contemplate losing. Science will undoubtedly play a complementary role.  After all, lunar science will  open up a visionary and inspiring landscape, hitherto completely unexplored by humanity,  while being relatively inexpensive. In what follows, I will summarize the science issues at stake, describe several of the proposed telescopes that are under detailed study,  continue with  discussion  of peripheral but key  strategic issues, and conclude with  suggestions for  proposed future actions.

\begin{figure}
\vspace{-2 cm}
\includegraphics[width=0.7\textwidth, angle=-90]
{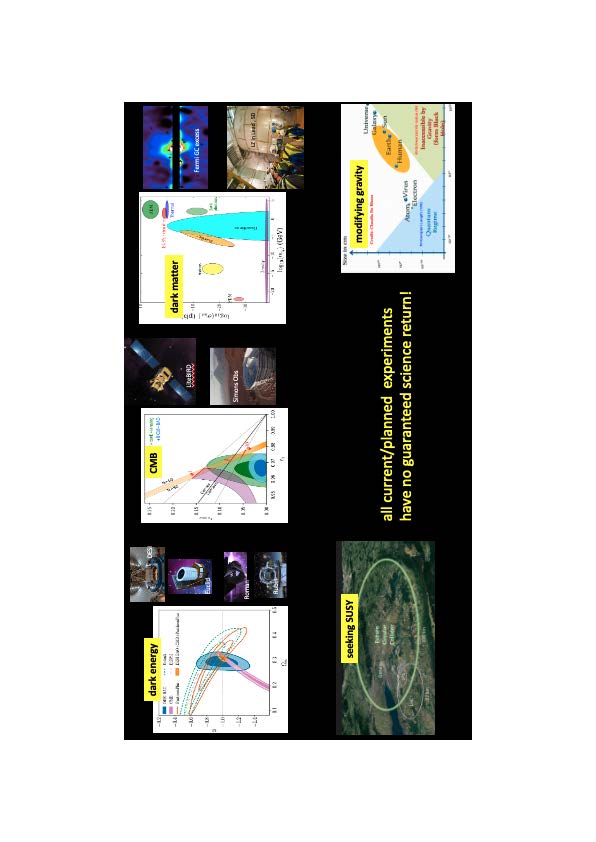}
\caption{Ongoing and proposed terrestrial and space experiments that probe dark energy, the primordial tensor to scalar ratio from inflation, direct and indirect  searches for dark matter signals, high energy collider experiments to probe creation of dark matter particles, and possible deviations from Einstein gravity. All  seek to explore our cosmic origins. Common to all is that there are no guaranteed science returns.}
\label{fig1}
\end{figure}

\section {Dark ages: the new frontier}
Cosmology seeks to go back in time to penetrate the obscurity of the past. Our current targets are the first stars  and the first galaxies, and we are making steady progress to explore these goals. 
However the real frontier lies further back in time. It spans the dark ages before any structures had formed, in the   first million years after the Big Bang, and begins around the end of the epoch of inflation in the first $10^{-36}$ second of time when the quantum seeds of all future structure  were generated. 

The dark ages begin at $z\sim 100,$ when the hydrogen temperature first  falls below that of the CMB and heralds the onset of structure formation. 
There is an interesting analogy with the CMB. First, there was detection of the cosmic microwave background radiation, in 1964, culminating in the discovery of its isotropy by the COBE. satellite. 
 The monopole detection led to the measurement of the  high precision CMB blackbody spectrum by the COBE FIRAS spectrometer in 1990. This was a major step in confirming the model of the very early years, back to a few months after the Big Bang, and occurred a decade and a half  after the discovery of the CMB.

Then there was the search for the infinitesimal fluctuations. After the COBE differential radiometer map of large angular scale primordial fluctuations,  the quest for the small angular scale fluctuations that directly mapped the seeds of structure formation was to take another decade. The  culmination was the magnificent WMAP and PLANCK    all-sky maps by satellites   launched in 2001 and 2009, respectively. that   set a new paradigm for cosmology.  The CMB fluctuations  added unprecedented precision by improvements of an order of magnitude or more  to determination of cosmological parameters and cosmological structure evolution. 

The dark ages 21 cm searches will surely follow a similar path. The first goal is to  detect the monopole component that signifies the onset of the dark ages before any galaxies formed, as the    baryonic component cools relative to the background radiation while and after thermal decoupling occurs due to the neutralization of the Universe. The sequel will be to detect the fluctuations in the baryon content that map  the onset of structure formation.
The long term goal will be to use the immense   amount of information in hydrogen maps of the dark ages  to provide a unique and guaranteed probe of inflation.  Having a  guaranteed science return was a crucial factor in funding the many CMB  experiments, and similar logic will surely apply to exploration of the dark ages.

%
%
%
%

\subsection{21cm monopole}
After the CMB decoupled  at  $z\sim 1000,$ the baryons and radiation remain coupled by thomson scattering via the residual electrons. As the density falls with the expansion, the coupling weakens and by $z\sim 100,$ the baryons  have thermally decoupled and the baryon temperature $T_b$  falls below that of the CMB temperature $T_\gamma.$  In absorption, at the  21cm line frequency 1420 MHz redshifted into the range $z= 30-100$ to frequency $47-14$ MHz, the differential hydrogen brightness temperature $T_b$ relative to the CMB amounts to \cite{pritchard2010}
$$
T_b=60{\rm mK}
\left(
 \Omega_b h\over {0.033} \right)
 \left(
 \Omega_m h\over {0.27}
\right)^{-1/2} x_{HI}\left(T_s-T_{\gamma}\over T_s\right)
\sqrt{1+z\over 50}.
$$
Here $x_{HI}$ is the (predominantly) neutral fraction, $T_s$ is the hydrogen spin temperature, $ h$ is the normalised Hubble parameter , and $\Omega_b$ and $\Omega_m$ are the normalised baryon and dark matter densities.

\begin{figure}
\vspace{1 cm}
\includegraphics[width=0.9\textwidth]
{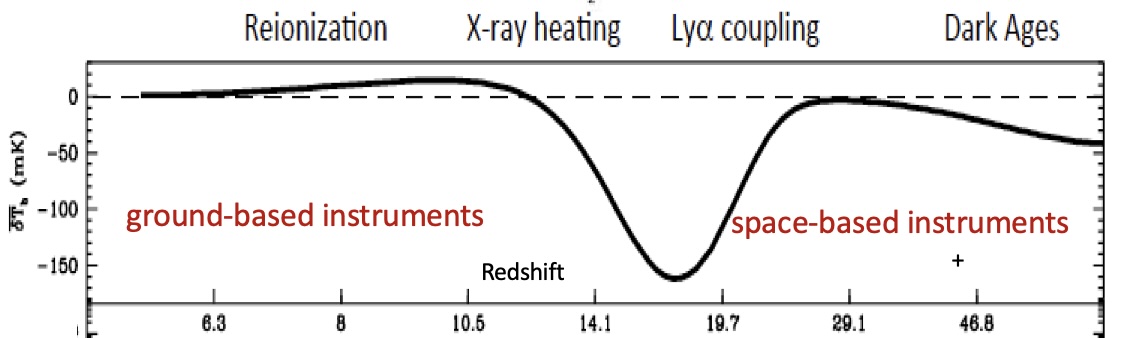}

\caption{The dark ages shadow, after \cite{pritchard2010}, is shown as the 21cm brightness temperature in mK.   Ground-based experiments probe the Universe to redshift of about 20, but are confronted by astrophysical sources of ionizing radiation that prevent any detection of the dark ages signal. One has to go to space experiments, and ultimately to the lunar far side, to optimize any possibility of detecting hydrogen from the dark ages  in absorption against the CMB in the   redshift  range $   30 \ltsim z \ltsim 100 $ at radio  frequencies   $\sim 50-10$MHz . }

\label{fig2}
\end{figure}

In the redshift range $30\ltsim z\ltsim 100$, there is little competition from ambient radiation fields, but  as structure forms, at lower redshift, the spin temperature recouples to the baryon temperature  by Lyman alpha coupling  and  x-ray heating of the IGM ($z\sim 30  \gtsim 10$) and IGM reionization ($z\ltsim10$). Hence the 21cm dark ages signal can be detectable at redshifted 21 cm frequencies in the range $\sim 10 - 50 $ MHz provided we can attain the required sensitivity. We seek a tens of mK signal in the presence of  a foreground of diffuse synchrotron emission from the MWG of the order of thousands of degrees Kelvin \cite{rogers2008}.

This will be a difficult measurement, and it is impossible from the Earth  to achieve such sensitivity (Figure 2). This is  because of  low frequency terrestrial radio interference and especially noise at low radio frequencies from the ionosphere.  The most radio-quiet environment in the inner solar system is on the far side of the Moon, which is of course shielded from terrestrial radio interference. Several  lunar experiments are being designed for probing the dark ages, and some are even under construction,  beginning with the first goal: detection.

\subsection {Detection of the dark ages}
The first far-side radio experiment Chang'e 4  landed in 2019 but was blighted by local radio interference from the lander  that limited its scope for studying regolith radio scattering properties.Three key experiments designed to ultimately detect the monopole signal of the dark ages have  launch dates projected to be within the next few years.
LuSEE-Night (for  LUnar Surface Electromagnetic Explorer)  is a NASA CLPS mission led by Stuart Bale (UCB) and is a low radio frequency crossed-dipole system (0.1-50 MHz) with 3m long wire antennae.  LuSEE-Night will be launched on a Firefly rocket to land on the lunar dark side in November 2026 and begin by measuring the low frequency radio foregrounds and regolith properties, the key obstacles to obtaining the dark ages signal from the early Universe. If battery power endures over many lunar nights, with recharges in the lunar day,  LuSEE-Night may just be able to achieve detection of the weak dark ages signal, although it is expected that the principal outcome will be an improved understanding of the role of lunar regolith in scattering low frequency radio waves. It is
very likely that  next generation version swould be needed for a definitive measurement. of the dark ages signal. These possibilities are described below.

There is competition from India and China for detection of the dark ages signal. PRATUSH (Probing ReionizATion of the Universe using Signal from Hydrogen) is a two year mission of an orbiting  dipole antenna and receiver that seeks to probe the dark ages over  40-200 MHz. It is expected to be launched within the next 5 years by ISRO into a circumlunar orbit. From China,   Hongmeng, also known as 
DSL (for Discovering the Sky at the Longest wavelengths), is  a lunar orbiting constellation of one mothership (to send data back to Earth) and nine daughter nano-satellites (to take data interferometrically on the distant universe from above the lunar far side) proposed by Xuelei Chen (NAOC). Hongmeng is expected  to be launched by CNSA to orbit the Moon at 300 km altitude  with sensitivity over 30-120 MHz. The experiment has been approved  by the CAS New Horizons program,  and is being prepared for launch  in 2028.

\begin{figure}
\vspace{-3 cm}
\includegraphics[width=0.8\textwidth,  angle=-90]
{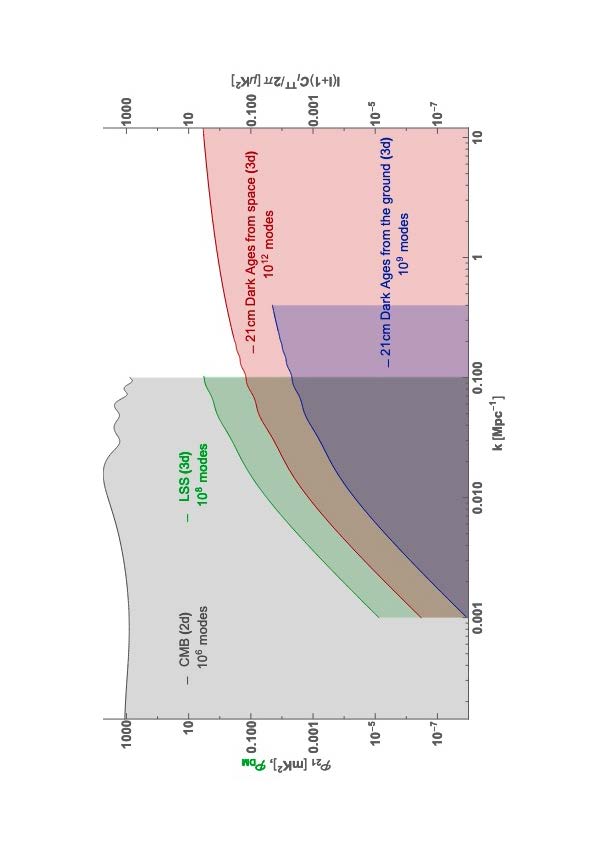}

\caption{The power in the 21cm fluctuations: trillions of modes \cite{cole2022}. 
I show several probes of the primordial fluctuation power spectrum. Note the y-axes are different for each probe. In grey is the TT
angular power spectrum in units of $\mu\rm K^2$  as shown on the right-hand axis, with multipoles roughly
mapped to wavenumbers by $\ell \sim 14000k/\rm Mpc^{-1}$. In green is the dimensionless 3d matter power
spectrum LSS computed with CAMB at redshift 1, where large-scale structure probes such as
LSST and EUCLID will be sensitive  on scales between $k \sim 0.001 - 0.1 \rm Mpc^{-1}$  up to around
redshift 2.5. In blue is the 3d 21cm power spectrum  in units of mK$^2$ at redshift 27, which is
the highest redshift accessible from ground-based experiments such as HERA and SKA. In red is
the 3d 21cm power spectrum in units of mK$^2 $ at redshift 50, which would be accessible from the
Moon. Note that the maximum k for 21cm experiments depicted here  is solely based on the angular resolution
for adopted maximum baselines.}

\label{fig3}
\end{figure}

\subsection{Next step: the fluctuations} 
Once there is  detection of the monopole absorption signal of the dark ages, the search will begin in earnest for   the inevitable angular fluctuations. This is a guaranteed signal, as was the case with the CMB, but it will be far more difficult because of spatially varying foregrounds.

\subsubsection{Probing inflation with the CMB: the gravity wave signal.}
First,  the good news. With the CMB,  we   have detectability of millions of pixels or independent modes in the data. These bits of information are equivalent to pixels on the sky, and the number is limited by radiative diffusive damping on small angular scales.
This limits the cosmological precision, once all competing foregrounds have been removed. The  detection of the scalar modes enabled WMAP and PLANCK  to reconstitute the primordial density fluctuations that generated cosmological structure.  Because of  damping,  the smallest scales we can reconstruct correspond to the predecessors of galaxy clusters and the most massive galaxies.  As we have seen,  this sufficed to usher in a new era of precision cosmology. The next CMB challenge is to detect a signature of inflation.

Huge efforts are currently underway to clean the CMB fluctuation maps signal and detect the tensor mode generated by the background of gravitational waves generated at the  end of inflation. Detection would verify that inflation occurred.
Unfortunately, the tensor mode prediction is not robust. There is no guarantee that the amplitude will be in the observable range, $T/S \gtsim 10^{-4}, $  for experiments such as that under construction by the Simons Observatory in Atacama. As mentioned above,  there are too many uncertainties in the inflationary framework, most notably  reheating temperature,  to guarantee an observable signal, quite apart from the need to remove  tensor foregrounds induced by both dust and especially by gravitational lensing,  for other than a  very restricted set of inflationary models.

\subsubsection{The ultimate goal.}
The Planck satellite all-sky maps demonstrated that the CMB is highly gaussian. Of course,  many foregrounds unavoidably  add non-gaussianity. These include dust and gravitational lensing as well as foreground emission, but the maps can be effectively cleaned using techniques with multifrequency and polarization information. 
 
  For this, we must go beyond the CMB and LSS, where the predicted limits  from generic inflation models are generally too low to be  attainable,  for example in terms of a primordial shear fluctuation signal characteristic of tensor modes. It is anticipated that improved spectral measurements of the primordial density fluctuation power spectrum may tighten these limits because of the immense increase in the number of modes on the sky once the 21 cm signal from the dark ages  can be probed \cite{cole2022},  as summarized in Figure 3.  Nevertheless,  brute force spectral precision is not the final answer, interesting though it may suffice  for insights on structure formation from the smallest scales.

Real progress will require more than improved spectral measurements. With enough precision, signatures of primordial non-gaussianity are  generically unavoidable.  One  has to go beyond the primordial bispectrum. Indeed there is general consensus that primordial non-gaussianity represents the new, hitherto almost completely unexplored, El Dorado of cosmology.

Recall first that primordial  non-gaussianity is the amplitude of the quadratic correction to the potential of a Gaussian random field, in this case  that of the CMB temperature fluctuations. It is 
parametrized as $\delta T/T(1+f_{NL}\delta T/T), $ where $\delta T/T$ represents the observed primordial  rms temperature fluctuations, measured to be gaussian-distributed to a high approximation.  To seek the primordial deviations  from gaussianity, much cleaning (dust, delensing, foreground contamination) of the  data by orders of magnitude is involved,  and it is non-trivial to eliminate the non-gaussianities  inevitably introduced by the cleaning algorithms.
Current  and envisaged limits are  from CMB $f_{NL} \ltsim  10$ \cite{planck2020}
and from LSS $f_{NL} \ltsim  1$ \cite{maartens2023}.
However this is simply not enough to make any significant progress in probing inflation. One needs far more modes.

In terms of information content,  these limits contain respectively $N\sim 10^6$, the diffusion limit,  and $N\sim 10^8-10^9$, the number of massive galaxies, that determine the number of available modes in the sky.  The Planck polarization data has even enabled us to detect tensor mode fluctuations that are generated by dust and by gravitational lensing. Measuring the highly gaussian nature of the observed CMB temperature fluctuations is a great success. But tensor mode fluctuations cannot robustly unravel inflation, unless we are fortunate enough to make a detection. This seems unlikely from any Bayesian perspective given the vastness of the inflationary landscape. 

Armed with  potentially much higher precision, the only robust prediction of inflation becomes within reach through the 21cm  signal from the dark ages. Via the vast anticipated information content  of 
$N\gtsim 10^{12}$ modes, as we describe below, this new window on the early universe 
will eventually allow detection of  
primordial non-gaussianity. The 21 cm fluctuations will be  
 immensely more powerful in terms of precision cosmology than either the CMB or LSS.  

However the challenge is huge.
More specifically,  \cite{maldacena2003} originally predicted that in single field inflation, the  minimal non-gaussianity would be 
$f_{NL} = -5/12(n_s-1)\sim 0.01 $ for the observed scalar index range $n_s =0.9743\pm 0.0034$  (ACT, DR6). 
 Subsequent attempts to generalize this limit to multifield inflation models did not significantly modify this lower bound, at least for the generic case of local primordial nongauussianities
\cite{ cabass2017}, \cite{bartolo2022}.

Many inflation models predict higher levels of primordial nongaussianity. However 
in the most pessimistic situation, the predicted level, using the canonical parametrization of primordial nongaussianity,  is $f_{NL} \sim 0.01$ for single or multifield inflationary models. 
In principle, attaining this threshold is feasible via 21cm data with suitably ambitious lunar far side experiments, as we describe next. 

\subsubsection{21cm to the rescue.}
There are trillions of 21cm modes because the building blocks of galaxies are hydrogen clouds of $\sim 10^6 \rm M_\odot, $ and there are millions of these per typical galaxy. These, or at least their precursor density fluctuations,  are present throughout the dark ages and allow  of order  $\sim 1/\sqrt{N}$, for  N information bits  on the sky. With $N\sim 10^{12}$  modes at redshifted 21cm, one can attain, in principle, 
a thousand times more precision in determining cosmological parameters than via the CMB  or LSS.  
With the power of the 21cm fluctuations from the dark ages,  we anticipate that at least $N\sim 10^{12}$ modes and $f_{NL}$ determinations at the percent level become feasible.
One might even improve on this, especially if one can take advantage of multifrequency tomographic slicing in redshift \cite{munoz2015}. 
Further refinements are feasible using 21cm cross-correlations with CMB fluctuations \cite{orlando2023}.
Indeed one recent study \cite{bull2024} suggests that the detectable dark ages signal could be even lower than the cosmic variance limit in the case of a large lunar array (Figure 4).

\begin{figure}
\includegraphics[width=\textwidth]{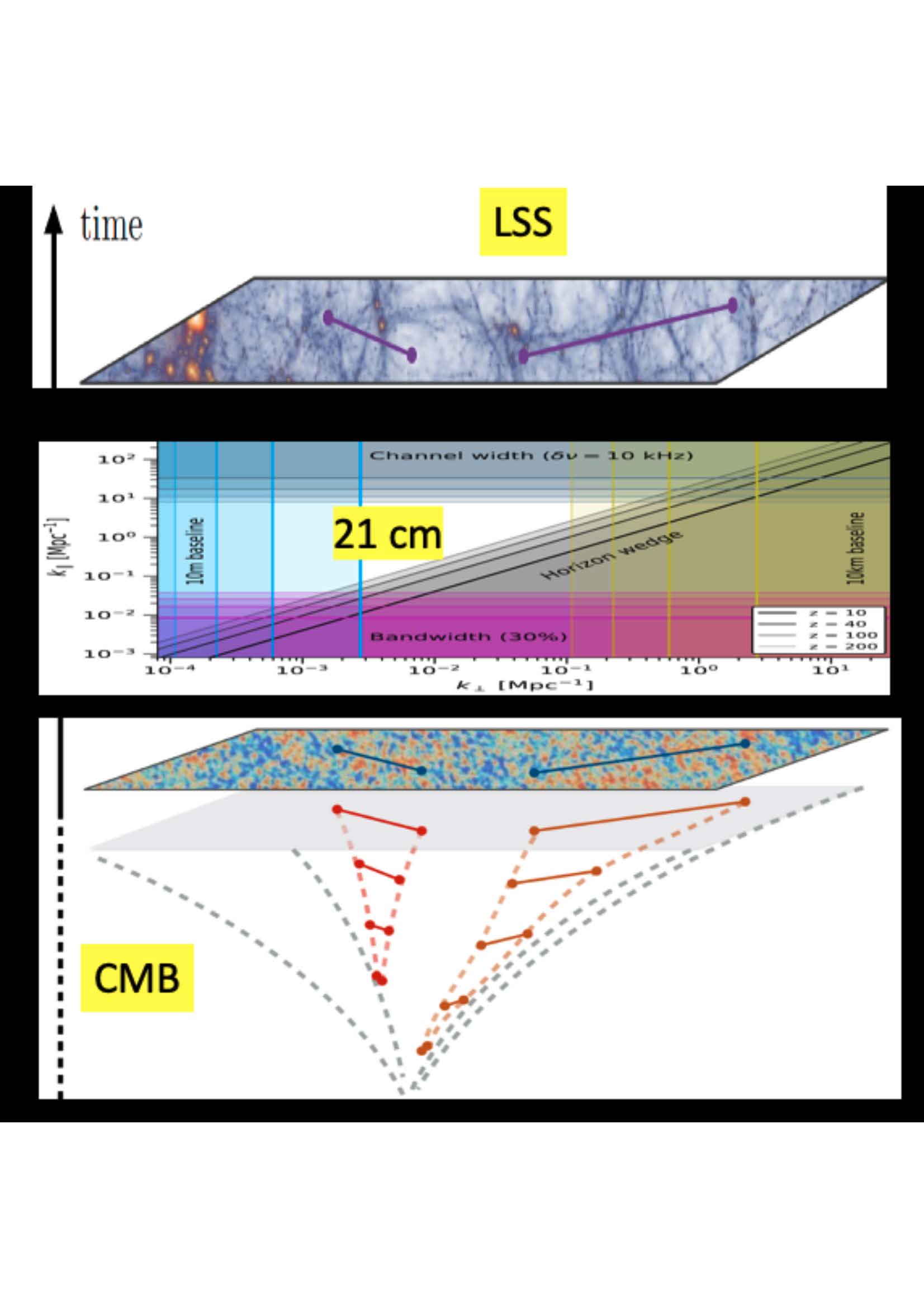}
\caption{The fluctuations at 21cm. Targets for a far-side lunar low frequency radio array, central panel taken from \cite{bull2024} with kind permission of the authors. 
The exclusion plot uses Fisher forecasting to show which cylindrical Fourier modes $k_{perp}, k_{par}$
are observable with a lunar  far side radio  interferometer for
four representative redshifts. Shaded regions are excluded from the observations due to various observational
effects: the foreground wedge (grey, diagonal); the frequency channel width (grey-blue, top); the fractional bandwidth of the
observation (magenta, bottom); and the minimum and maximum baseline lengths (blue, left and yellow, right respectively).
Representative numbers are assumed, with  10 kHz frequency channels, a 30\% effective bandwidth at each redshift,
and minimum and maximum baselines of 10m and 10km respectively. Other graphics: upper and lower panels are cartoon representations  of a cosmic collider   inspired by \cite{achucarro2022}. 
}
\label{fig4}
\end{figure}

\subsubsection{The future}
The dark ages  future lies with construction of large  lunar  far side radio interferometer arrays. 
Our target would be say 100 times the damping-limited CMB resolution, that is with a Fourier component reach of  $\ell\sim 10^5$ or wave number $k \sim 10 \,
\rm Mpc^{-1}.$
To achieve this angular resolution, the  optimal array baseline  should span a diameter  $\ell \lambda/2\pi$ or $d\sim$100 km. 
 One needs to fill this area and that requires of order ${{1}\over{2}} D^2/\lambda^2$ or $\sim 10^6$ dipoles.  
 The smallest scales probed would amount, with extensions to $\sim$ 3000 km  lunar  baselines,  to  $k \sim 250\,  \rm Mpc^{-1} $ \cite{decruijf2024}. That is an equivalent comoving  hydrogen cloud mass scale of $\sim 10^5 \rm M_\odot.$
 
In addition to the cosmology goal, there are inevitable  sources of injected energy, and our standard model 
predicts that there must be small deviations from a perfect blackbody spectrum. This would similarlly unleash trillions of modes, this time in blackbody radiative power, and have  implications for a complementary FIR spectrometer as we note below. 

Several pioneering  far side  surface arrays are under design.  The time frame for construction  is 2030-2050. They range from  FARSIDE with 128 antennae, potentially a NASA Probe mission,  to ESA's  ALO (Astrophysical Lunar  Observatory) with several hundred antennae imprinted on inflatable tubes,  to be delivered by the Argonaut Lander after 2030. China is planning LARAF (Large-scale array for Radio Astronomy on the Farside)  with 270 antennae over a 10 km baseline, and more ambitiously,  a 7200 element lunar far side  array  sensitive over 0.1-30 MHz with butterfly-shaped wire  antennae  deployed by astronauts over a 30 km diameter by 2035.
The largest interferometer array  planned to date is NASA's FarView, over a similar area and involving in situ manufacture of $10^5$ antennae (5-40 MHz), again with astronaut deployment and even further into the next decades.

In the analogous CMB fluctuation history, both large and small dishes are playing a key role in  the Simons Observatory  all-Chile configurations in obtaining the ultimate precision on foreground removal that is required by the science goals. Much  larger dishes would likely be essential in the low radio frequency  lunar  experiments in order  to complement the interferometric arrays and  overcome similar obstacles in obtaining a cleaned, foreground-removed, sky. 

One of  the most ambitious projects under consideration   by Nasa's JPL is the  LCRT (Lunar Crater Radio Telescope), a monolithic Arecibo-style low frequency radio aperture-filled low frequency radio telescope in a dark-side lunar crater  that spans kilometers. The parabolic  surface reflector of deployable wire mesh with  diameter of several hundreds of meters or  more.is embedded in the crater bowl. A  receiver in the focal plane is  suspended by cables spanning  the crater rim.

\section {Origin of the cosmic microwave background}

The CMB was created during the first months of the Big Bang, that is when the Universe was dense enough to guarantee complete thermalization of all  radiation sources.  A perfect blackbody spectrum was generated that is  characterized by a blackbody temperature  of 
$2.728 \pm 0.004 \rm K \, (95\% CL)$
 The intensity was precisely measured by COBE/FIRAS more than three decades
ago, with any deviations limited to a few  parts in $10^5$
\cite{fixsen1996}. 

After the first months, spectral distortions are inevitable that  encode  the subsequent thermal history of the Universe. Some are exotic (such as particle decays or even primordial black hole  evaporation). However  there are also inevitable distortions produced in the generally accepted  model of the early Universe.

 Our standard model of structure formation predicts that  throughout the radiation-dominated era, there must be small deviations from a perfect blackbody spectrum due to the radiation friction  between primordial adiabatic density fluctuations and the CMB on the scales of the precursors of dwarf galaxies.    This inevitably creates a $\mu $-type spectral distortion \cite{sunyaev1970} 
from  diffusive damping of the fluctuations\cite{hu1994}
 that is some four orders of magnitude below the FIRAS upper limit of $\mu < 9\times 10^{-5} (95\%CL)$
 \cite{fixsen1996}. Detection would provide a confirmation of the bottom-up CDM theory of structure formation. It is a guaranteed prediction of our standard  LCDM model.

 \begin{figure}
 \vspace{-1 cm}
\includegraphics[width=0.7\textwidth,angle=-90]{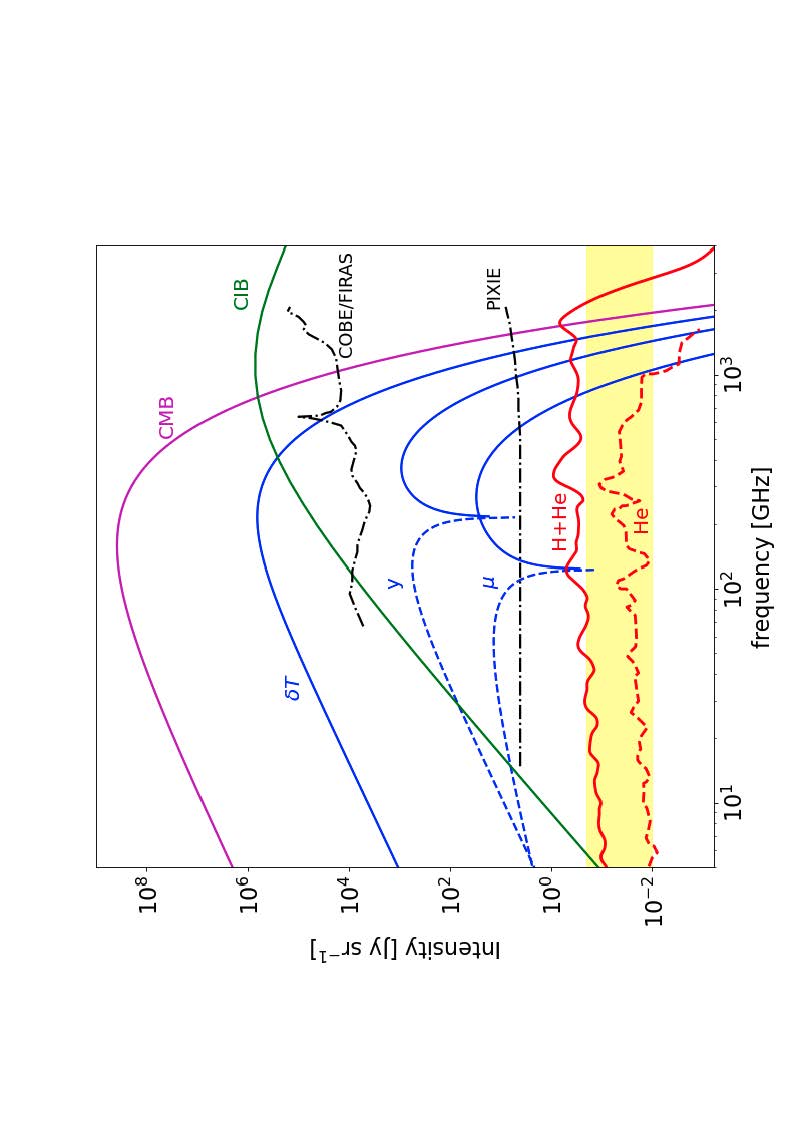}
\caption{CMB spectral distortion forecasts.
Reading down, CMB is purple line. temperature fluctuations are blue line, CIB is green line, FIRAS limits are black dashed line,  upper blue line (dashed plus continuous) is y prediction  in quadrature, lower  blue line (dashed plus continuous) is $\mu$ prediction  in quadrature, 
black dashed line is PIXIE 5 yr sensitivity, red lines are  hydrogen and helium recombination line predictions,
gold band is lunar goal and similar to proposed VOYAGE2050 range of sensitivity. Adapted from \cite{desjacques}.}

\label{fig5}
\vspace{-2 cm}
\end{figure}

 The best way to improve  on the COBE  FIRAS Michelson interferometer experiment that operated by differencing on an internal blackbody is with a Fourier Transform Spectrometer mounted on a space telescope. This has been  unsuccessfully  proposed  (in two calls) as a NASA  Explorer mission, with just a three order-of-magnitude improvement in sensitivity over COBE FIRAS \cite{kogut2024}, as well as  the unsuccessful PRISM mission in an ESA call  \cite{PRISM}.  
 
 A larger and more sensitive imaging FTS  is advocated by the ESA Voyage 2050 review as a primary  cosmology future L-class space mission \cite{chluba2021},   to be possibly selected in competition  with a gravitational wave interferometer \cite{esa}. However the proposed decadal cadence of ESA L-class missions leaves little prospect for selection of a cosmology mission prior to 2065.
 
 A lunar platform might be feasible for a CMB spectral distortion telescope..
 Such an alternative to a free flyer in the CMB case would be  a FTS-telescope combination mounted  in a  permanently dark and cold polar or near-polar lunar crater. The natural competition here is  with proposed gravity wave experiments, as discussed below. The crater FTS need not scan the sky, as lunar rotation would take care of this, and a significantly larger telescope is feasible compared to the possibilities of any conceivable free-flying L-class mission \cite{maillard2020}. 
 The advantage here  (Figure 5) is that by gaining an extra order of magnitude in sensitivity primarily via increasing  telescope aperture to a meter-class design, this would open the way to exploration of recombination line emission at and even before the last scattering of  CMB photons at $z\approx 1066,$ thanks to helium recombination at $z\sim 10^4$ \cite{hart2023}. This would be a remarkable breakthrough experiment for cosmology, providing the ultimate measurement of BBN-synthesised primordial helium and  demonstration that bread-and-butter astronomy operated  300,000 years after the Big Bang, as well as providing novel probes of fundamental physics.

 
 \section {Gravitational waves: bridging the gap}
 A major advance in astronomy has emerged with success of gravitational wave interferometers. These have  confirmed a major prediction of General Relativity, namely the existence of black holes, and have revolutionized our  understanding of stellar death and black hole mergers. With LIGO, VIRGO and KAGRA, we are now routinely detecting signals from black hole mergers and even neutron star-black hole mergers, enabling us to locate these events with unprecedented accuracy. 
 
  A new generation of gravitational wave experiments is under design with greatly  improved sensitivity. These include the ground-based Einstein and Cosmic Explorer  telescopes with $\sim 30$ km baselines,  which are expected to detect thousands of black hole mergers per day and see every astrophysical black hole merger out to $z\sim 20.$ The  LISA  (Laser Interferometer Space Antenna) gravitational wave space observatory consists of three  spacecraft  separated by millions of miles and trailing tens of millions of miles behind the Earth as we orbit the Sun. It is intended to detect  mergers of  massive black holes in AGN in the remote universe and clarify how supermassive black holes formed.  All three experiments are scheduled to operate from 2035. LIGO/VIRGO/KAGRA/EINSTEIN/COSMIC EXPLORER  operate at kHz to Hz frequencies, LISA at mHz frequencies. 
 
 This strategy does leave a decihertz frequency gap. 
 Bridging this gap from Hz to mHZ  is the goal of lunar gravitational wave observatories, and will allow us to follow the approach of black holes on longer time scales and lower frequencies before they actually merge. This is crucial for understanding the evolution of mergers. Lunar gravitational wave observatories are currently being designed to fill this frequency gap. The  extremely low level of seismic disturbances on the Moon allows precision seismology and interferometry. At least two lunar concepts are being studied \cite{cozzumbo2023}.
 
 
 The Moon is seismically remarkably quiet. The permanently shadowed polar craters provide ideal conditions for advanced lunar seismometry. One can span the vibrations of the Moon in the decihertz band at high sensitivity. That is the goal of LGWA, a proposed  lunar gravitational wave telescope, with  peak sensitivity near 1mHZ and an array of cryogenic  inertial sensors deployed in a permanently shadowed polar crater \cite{LGWA}. 
 
 Another and complementary concept uses  gravitational wave interferometry in the LVK tradition, the LILA (Laser Interferometer Lunar Antenna). This project  consists of three laser stations straddling a dark crater with a few km diameter, chosen for the ambient ultrahigh vacuum conditions \cite{jani}. The idea is to  achieve peak sensitivity in the subHz range.
 
 Strong competition for the decihertz range will be provided by a Chinese free flyer TianQin, a LISA-like configuration of three satellites in high earth orbit at some 100,000 km altitude and separation, to be flown in 2035 \cite{TQ}.

 \section {Exoplanets and optical interferometers}
Exoplanet research is one of the most active research areas in astronomy.  The ultimate goal is to answer one of the most profound questions ever posed by humanity: are we alone in the Universe?
First we need to find biotracers in exoplanet atmospheres to demonstrate that conditions for life are, hopefully, ubiquitous.

 Of course we have no idea about the link between  the necessary  ingredients, that might include habitable zones, oxygen-rich atmospheres, oceans,  forests  or other possible criteria of habitability,  and the eventual emergence of life.  Therefore the gauntlet is thrown down, to launch telescopes that can assess these issues. The leading candidate at present may be  NASA's Habitable Worlds Observatory, destined to launch in the mid 2040's.  Its announced goal is to study biosignatures in the nearest 25 earth-like exoplanets. Many astronomers worry that the achievable statistics  might not be sufficient to fulfill the ambitious announced goals.

The Moon can provide a promising environment for interferometry. The lack of atmosphere means that there are  ideal sites for optical and IR interferometry. Without atmospheric refraction, integration times for optical interferometers are in principle hours, as opposed to milliseconds for terrestrial optical interferometers. This means one does not even need extremely large telescopes. A widely spaced array of small or medium-sized telescopes is a promising option to achieve unprecedented resolution and biosignature searches.  

One frequently voiced objection to lunar telescopes  is that of the pervasiveness of lunar dust,  as encountered by the Apollo astronauts. However dust  is charged, and experts believe that the dust probably photolevitates  in the day but falls to the surface at night. 
This may explain in part why a small  telescope operating on the Moon has taken many   images of stars and galaxies.  
The 15-centimetre LUT (Lunar-based Ultraviolet Telescope)  telescope   on board  the Chang'e 3 lander operated flawlessly for at least 5 years from  2013. Location on  $\sim 2 m   $ towers  would even surmount the likely dust  daytime scale-height.

\subsubsection{Projects under consideration.} Ambitious lunar interferometry projects are under development.
MoonLITE is a two-element optical interferometer with a 100m baseline and 5 cm diameter telescopes that is under consideration \cite{vanbelle2024}. Integration times vastly exceed those of  terrestrial interferometers, allowing high sensitivity for small apertures.  This is expected to provide  a pioneering format for far more ambitious lunar interferometers. These  would include larger telescopes, longer  baselines and larger element apertures. 

The next step could be the AeSI (Artemis-enabled Stellar Interferometer). With a baseline of up to 0.5 km, this proposed UV/optical interferometer design could initially accommodate six or more elements of meter aperture along with delay line rovers and beam combiners in the 9mx18m fairing and cargo down mass of 100 metric tons of a SpaceX Starship Human Landing System or of other similar lunar landers under study \cite{rau2024}. Remarkable science could be attainable at a resolution of order 0.1mas or better. 

Of course, one  ideally  needs much higher resolution to search for biofeatures even on the nearest exoplanets. 
Looking ahead, and this may take several decades, will be an ambitious plan to push optical interferometry on the Moon to towards $\mu$as resolution (Figure 6). This is the science breakthrough point: we could anticipate imaging  the nearest exoplanets.  

One example is the planned hyperscope. This  consists of a string of $\sim$ 5 m dishes on pivots in a polar shadowed crater,  filling a hemispherical bowl with the focus array, beam combiner and appropriate time delay lines supported by cables strung from the crater rims \cite{schneider2024}.

This would approach  the holy grail of exoplanet research. Imaging the nearest exoplanets from the Proxima Centauri to the Trappist 1 planetary systems at $\mu$as resolution could answer some our wildest speculations about conditions for (and even existence of)  life elsewhere in the Universe. Only the Moon could provide the necessary observational platform conditions for such a visionary endeavour.

\subsubsection{FIR telescope.}
One needs to go beyond interferometers of course. Monolithic telescopes with their wider fields of view will surely be a lunar target, especially in the infrared  and for larger telescopes than we can envisage from the spectroscopic, size and budgetary limitations of  terrestrial or even   free flyer designs. We can go far deeper than JWST with, say, a 13 m aperture IR telescope located in a permanently shadowed polar crater, whose mirror segments fit into the bearings of projected launch vehicles \cite{maillard2024}. Looking further ahead, we can imagine tapping  advantages of  low gravity and lack of atmosphere of the Moon to build an optical  telescope of 100m aperture.  The science return will range from novel spectroscopic signatures of exoplanets to detection of the first stars and massive black holes in the Universe, among the most ambitious goals of cosmology and astronomy.

\vspace{1 cm}
\begin{figure}
\vspace{-2 cm}
\includegraphics[width=0.7\textwidth, angle=-90]{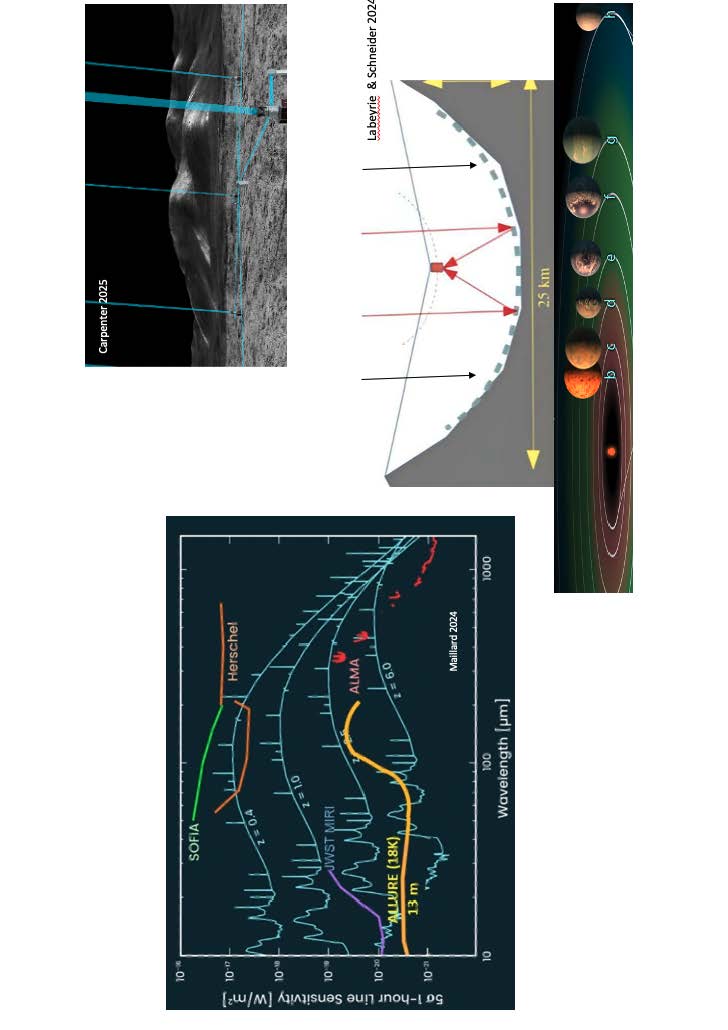}
\caption{From IR telescopes to optical interferometry. The projected sensitivity is shown for a 13 m IR telescope, with aperture chosen so that segmented design fits in the space craft bearings, and takes full advantage of being sited on the floor of a permanently shadowed polar crater \cite{maillard2024}.  The interferometry projects include an Artemis-enabled optical interferometer array with a baseline of up to 500m 
\cite{rau2024}. The ultimate plan for a lunar interferometer is a hyperscope, an Arecibo-type design that sites an arc of 100 5m telescopes in  the basin of a permanently shadowed lunar polar crater. Delay lines are  incorporated into the focal plane that is itself suspended from rim-spanning cables,  and capable of imaging at $\mu$sec resolution \cite{schneider2024},  as in an  artist's conception, shown here (courtesy A. Labeyrie and J. Schneider)  of the Trappist planetary system. }
\label{fig6}

\end{figure}


 \section{Final thoughts}
\subsection {Are lunar telescopes too expensive?}
A first reaction to many of these telescope designs is to ask how  the funding will be provided.
This should not be a problem given that the infrastructure, including launch,   delivery, communications and even deployment,  will largely be covered by the ARTEMIS program, if it goes ahead as planned, despite its most recently announced delay. However a first reaction  is initially  negative: a significant science overhead is mostly not budgeted, and would be vulnerable to future constraints even if  included.   However, other countries, most notably China, might enthusiastically fill this gap.

The history of space telescopes tells another story. Apollo was hugely expensive, but the funding was mostly transient. The budget peaked over five years, then returned to a steady state of some 5\% of the annual NASA budget. The total cost of Apollo in today's dollars was some  \$200bn, to land 12 men on the Moon. Nasa funding peaked at some 5\% of government spending, whereas today it is a tenth of that, or about \$25bn in FY2025, although significantly  less is projected for FY2026. 

However looking on the bright side, the Apollo success eventually seeded the development of the space shuttle  some two decades later.   The space shuttle program cost some \$300bn. 
This was used to launch major telescopes such as Hubble, built for some \$2bn. Including running costs,  that perhaps quadruple this figure, we see that  the key science project cost only a few percent of the launcher costs.
Science was relatively cheap. And generated a bonus that inspired humanity.

The same logic should apply today.  Deep space exploration consumes a third of the current NASA budget. And the estimated cost of ARTEMIS to date is around \$100bn, with substantial increments for  crewed   landings still to come in 2027 or later. The NASA budget remarkably appears to remain intact for the forthcoming ARTEMIS missions, thanks largely  to the space race with China. A few percent of infrastructure costs, just a minor fraction of the lunar budget,  would suffice to fund almost all of the lunar  telescopes described above. Let us hope our space agencies can be convinced to look to the far horizon to invest in human inspiration along with technology and commercial goals. 

Fortunately, all space agencies  agree that science, and most specifically the science led by space exploration,  sets inspiration for the entire human race. And it is inevitable that the international space race will have winners and losers.
The healthy spirit of competition should ensure that lunar exploitation, and specifically its science ambitions,   set the targets for the benefit of humanity.

\subsection {Competition}
The space race to the Moon is underway, with lead players  the US, China, and Europe,  and significant supporting roles being played by India, Russia and Japan. However coordination is lacking. The Outer Space Treaty was signed at the UN in 1967 and has some  115 signatories. It sets out 
the basic framework on international space law,  with such highlights as  forbidding national appropriation of lunar sites,  exclusion of using nuclear weapons in outer space, and avoidance of pollution. However no provision was made for enforcement. There is no effective system for settling disputes,

 NASA  plans to send  astronauts to the Moon in 2027 and Chinese astronauts are expected to land  by 2030.  There are plans for exploitation of lunar resources, including mining of ice in polar craters for use in developing fuel reservoirs. These are important  for future travel to Mars and more generally for solar system exploration, as well as for developing local biospheres. Mining for rare earth elements, and most futuristically He$^3,$ relatively abundant on the Moon, will be a source of intense commercial activity. 
 
 Science can  benefit from the competition. 
 The space race includes telescope projects, most notably for developing low frequency radio observatories on the lunar far side.
 Other projects include telescopes  in permanently shadowed polar craters and interferometers that study nearby exoplanets with unprecedented precision.
 
 There are relatively few  suitable sites for our telescopes. We will need to develop some sort of allocation system that will preserve certain areas for science, others  to be  designated for commercial exploitation. We need to avoid the Wild West scenario. In any free-for-all, science would be the first loser.  Rather the exploitation of Antarctica should serve as a guide to develop a future template. Unfortunately the lunar stakes are far higher. 
 
 We are already starting the necessary discussions by developing guidance for optical and radio frequency electromagnetic wave pollution, fundamental for lunar astronomy. That is the very minimum requirement for astronomers.
 
  However thats only the beginning of a tortuous process.
 The UN Committee on the Peaceful Uses of Outer Space (COPUOS) was set up in 1959 and is currently active in this field.
 However there  are  vast obstacles to circumvent, most notably surrounding choice of protected sites for lunar telescopes, the interface with exploitation of lunar mineral resources and control of lunar debris from human activities. Limited numbers of polar sites are available, and there will be strong competition for lunar resources. The interplay between political, commercial and scientific pressures remains to be resolved.

\subsection {What next?}
Two of the  most profound questions ever posed in science can be addressed by lunar telescopes. How  did the Universe begin? Are we alone  in the Universe? The unique and accessible lunar environment is an essential part of our future. Pollution is a critical issue, and for astronomers this means controlling  light and radio wave backgrounds. And this requires coordination with exploration and commercial activities

Let us focus on  the science. Most immediately, one has to begin with pilot projects  that are demonstrators to set the scene for more grandiose telescopes. 
These cosmic explorers are inexpensive but compete with more practical experiments designed to study, for example, lunar geology and space weather. One has to achieve a healthy balance.

 One first example highlighted by the cosmology community is  a simple radio dipole-type  antenna on the lunar far side or in orbit around the moon to detect  the monopole dark ages signal. Even here there are  environmental backgrounds that first need to be understood, most notably to do with the radio wave reflectivity of the lunar regolith. 
 
 Several  telescope projects are underway by different countries. Planned low radio frequency far side telescopes  include the imminent deployments of LUSEE-Night on the far side (USA), followed by  Hongmeng  (China) and PRATUSH (India) in orbit over the far side within the next five years. 
 
 The follow-up on decadal timescales would include dipole arrays of increasing size and complexity, perhaps inevitably culminating with in situ construction as for the FarView radio interferometer.

Another compelling  example pioneered by  the exoplanet community is an optical interferometer. The pilot project could be suitcase-size as the case for CLPS-scale projects such as 
MoonLITE. The sequels aim at increasing baselines and larger mirrors, in association with deployment by astronauts, with the ultimate goal of attaining  $\mu$as resolution.  

The immediate lesson for astronomers is that we need to lobby our space agencies to fund the pilot projects. They are relatively  inexpensive. But success will create the momentum to go forward to the ultimate dream of scaled-up experiments that will attack our  biggest questions about the Universe. How did the Universe begin? Are we alone?

\subsubsection {Acknowledgements.}
I thank my colleagues Philip Bull,  Jens Chluba, Philippa Cole,   Vincent Desjacques and Jean-Pierre Maillard for invaluable discussions and help in preparation of the figures.

\end{document}